\documentclass[12pt]{article}

\usepackage{named}
\usepackage{graphicx}
\usepackage{url}

\newcommand{\idr}{\textsf{idR}}
\newcommand{\idid}{\textsf{idID}}

\newcommand{\tid}{\textsf{TID}}

\newcommand{\card}[1]{|{#1}|}

\newcommand{\local}{\textsc{EasyLocal++}}

\begin{document}

\title{Local Search Techniques for Constrained Portfolio Selection Problems}

\author{Andrea Schaerf\\
Dipartimento di Ingegneria Elettrica, Gestionale e Meccanica,\\
Universit\`a di Udine,\\ 
via delle Scienze 208, I-33100,\\
Udine, Italy,\\
email: \texttt{schaerf@uniud.it}}

\maketitle


\begin{abstract}
  We consider the problem of selecting a portfolio of assets that
  provides the investor a suitable balance of expected return and
  risk. With respect to the seminal \emph{mean-variance} model of
  Markowitz, we consider additional constraints on the cardinality of
  the portfolio and on the quantity of individual shares.  Such
  constraints better capture the real-world trading system, but
  make the problem more difficult to be solved with exact methods.
  
  We explore the use of local search techniques, mainly tabu search,
  for the portfolio selection problem. We compare and combine previous
  work on portfolio selection that makes use of the local search
  approach and we propose new algorithms that combine different
  neighborhood relations. In addition, we show how the use of
  randomization and of a simple form of adaptiveness simplifies the
  setting of a large number of critical parameters. Finally, we show
  how our techniques perform on public benchmarks.
\end{abstract}

\section{Introduction}
\label{sec:introduction}

The \emph{portfolio selection} problem consists in selecting a
portfolio of \emph{assets} (or \emph{securities}) that provides the
investor a given expected return and minimizes the \emph{risk}.  One
of the main contributions on this problem is the seminal work by
\citeauthor{Mark52} (\citeyear{Mark52}), 
who introduced the so-called
\emph{mean-variance} model, which takes the variance of the portfolio
as the measure of risk.  According to Markowitz, the portfolio
selection problem can be formulated as an optimization problem over
real-valued variables with a quadratic objective function and linear
constraints (see \cite{Mark59} for an introductory presentation).

The basic Markowitz' model has been modified in the recent literature
in various directions. First, \citeauthor{KoYa91} (\citeyear{KoYa91})
propose a linear versions of the objective function, so as to make the
problem easier to be solved using available software tools, such as the
simplex method. On the other hand, with the aim of better capturing
the intuitive notion of risk, \citeauthor{KoSu95} (\citeyear{KoSu95})
and \citeauthor{MTXY93} (\citeyear{MTXY93}) studied more complex
objective functions, based on the notions of \emph{skewness} and
\emph{semi-variance}, respectively.  Furthermore, several new
constraints have been proposed, in order to make the basic formulation
more adherent to the real world trading mechanisms. 

Among others, there are constraints on the maximal cardinality of the
portfolio \cite{CMBS00,Bien96} and on the minimum size of trading lots
\cite{MaSp99}. Finally, \citeauthor{Yosh96} (\citeyear{Yosh96}) and
\citeauthor{GlMH96} (\citeyear{GlMH96}) consider multi-period
portfolio evolution with transaction costs.

In this paper we consider the basic objective function introduced by
Markowitz, and we take into account two important additional
constraints, namely the \emph{cardinality} constraint and the
\emph{quantity} constraint, which limit the number of assets and the
minimal and maximal shares of each individual asset in the portfolio,
respectively. 

\vspace{\baselineskip}

The use of local search techniques for the portfolio selection problem
has been proposed by \citeauthor{Roll97} (\citeyear{Roll97}) and
\citeauthor{CMBS00} (\citeyear{CMBS00}). In this paper, we depart from
the above two works, and we try to improve their techniques in various
ways.  First, we propose a broader set of possible neighborhood
relations and search techniques. Second, we provide a deeper analysis
on the effects of the parameter settings and employ adaptive
evolution schemes for the parameters.  Finally, we show how the
interleaving of different neighborhood relations and different search
techniques can improve the overall performances.

We test our techniques on the benchmarks proposed by \citeauthor{CMBS00},
which come from real stock markets.  

The paper is organized as follows.
Section~\ref{sec:portfolio-selection} introduces the portfolio
selection problem and its variants. Section~\ref{sec:local-search}
recalls the basic concepts of local search.
Section~\ref{sec:application} illustrates our application of local
search techniques to the portfolio selection problem.
Section~\ref{sec:experimental-results} show our experimental results.
Section~\ref{sec:related-work} discusses related work. Finally,
Section~\ref{sec:conclusions} proposes future work and draws some
conclusions.

\section{Portfolio Selection Problems}
\label{sec:portfolio-selection}

We introduce the portfolio select problem in stages. In
Section~\ref{sec:basic}, we present the basic unconstrained version of
Markowitz. In Section~\ref{sec:constraints}, we introduce the specific
constraints of our formulation. Other constraints considered in the
literature but not in this work are mentioned in Section~\ref{sec:extensions}.
\subsection{Basic formulation}
\label{sec:basic}


Given is a set of $n$ \emph{assets}, $A = \{a_1,\dots,a_n\}$.  Each
asset $a_i$ has associated a real-valued \emph{expected return} (per
period) $r_i$, and each pair of assets $\langle a_i,a_j\rangle$ has a
real-valued \emph{covariance} $\sigma_{ij}$.  The matrix
$\sigma_{n\times n}$ is symmetric and each diagonal element
$\sigma_{ii}$ represents the \emph{variance} of asset $a_i$.  A
positive value $R$ represents the desired expected return.

A portfolio is a set of real values $X = \{x_1,\ldots,x_n\}$ such that
each $x_i$ represents the fraction invested in the asset $a_i$. The
value $\sum_{i=1}^{n}\sum_{j=1}^{n} \sigma_{ij} x_{i} x_{j}$
represents the variance of the portfolio, and it is considered as the
measure of the risk associated with the portfolio. Consequently, the
problem is to minimize the overall variance, still ensuring the
expected return $R$.  The formulation of the basic (unconstrained)
problem is thus the following \cite{Mark59}.

\begin{eqnarray}
min  & \displaystyle \sum_{i=1}^{n}\sum_{j=1}^{n} \sigma_{ij} x_{i} x_{j} & \nonumber\\
s.t. &  \displaystyle\sum_{i=1}^{n} r_i x_i \geq R & \label{con:return}\\
     &  \displaystyle\sum_{i=1}^{n} x_i = 1 & \label{con:sum-one}\\
     & 0 \leq x_i \leq 1 & (i = 1,\dots,n) \label{con:limit}
\end{eqnarray}

This is a quadratic programming problem, and nowadays it can be solved
optimally using available tools\footnote{For example, an online portfolio
  selection solver is available at 
  \url{http://www-fp.mcs.anl.gov/otc/Guide/CaseStudies/port/}}
despite the NP-completeness of the underlying decision problem.

\begin{figure}[htbp]
  \begin{center}
    \includegraphics[width=90mm,angle=-90]{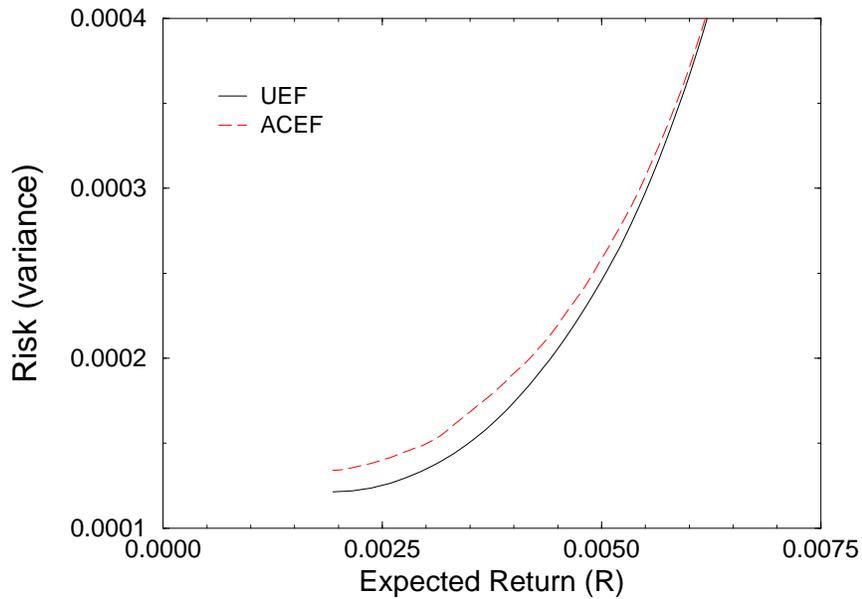}
    \caption{UEF and ACEF for instance no.\ 4}
    \label{fig:uef+acef}
  \end{center}
\end{figure}

By solving the problem as a function of $R$, we obtain the so-called
\emph{unconstrained efficiency frontier} (UEF), that gives for each
expected return the minimum associated risk. The UEF for one of the
benchmark problems of \citeauthor{CMBS00} (\citeyear{CMBS00}) is
provided in Figure~\ref{fig:uef+acef} (solid line).

\subsection{Additional constraints}
\label{sec:constraints}

In our formulation, we consider the following two additional 
constraint types:

\begin{description}
\item[Cardinality constraint:] The number of assets that compose
  the portfolio is limited. That is, a value $k \leq n$ is given 
  such that the number of $i$'s for which $x_i > 0$ is at most $k$.
\item[Quantity constraints:] The quantity of each asset $i$ that is
  included in the portfolio is limited within a given interval.
  Specifically, a minimum $\epsilon_i$ and a maximum $\delta_i$ for
  each asset $i$ are given, and we impose that either $x_i = 0$ or
  $\epsilon_i \leq x_i\leq \delta_i$.
\end{description}

These two constraint types can be modeled by adding $n$ \emph{binary}
variables $z_1,\dots,z_n$ and the following constraints.

\begin{eqnarray}
  \label{eq:cardinality}
  & \displaystyle \sum_{i=1}^{n} z_i \leq k & \label{con:card}\\
  & \epsilon_i z_i \leq x_i \leq \delta_i z_i & (i = 1,\dots,n) \label{con:mm_limit} \\
  & z_i \in \{0,1\}& 
\end{eqnarray}

The variable $z_i$ equals to $1$ if asset $a_i$ is included in the
portfolio, $z_i = 0$ otherwise. The resulting problem is now a mixed
integer programming problem, and is it much harder to be solved using
conventional techniques.

We call CEF the analogous of the UEF for the constrained problem.
Given that we do not solve the problem with an exact method, we do not
actually compute the CEF, but what we call the ACEF (\emph{approximate
  constrained efficiency frontier}).  Figure~\ref{fig:uef+acef} shows
the ACEF (dashed line) we computed for the same instance for the
values $\epsilon_i = 0.01$, $\delta_i = 1$ (for $i=1,\dots,n$), and $k
= 10$.

Notice that when the return is high, the distance to UEF is very small
because typically large quantities of a few assets are used, and thus
Constraints~(\ref{con:card}) and~(\ref{con:mm_limit}) don't come into
play. 

In particular, in the instance of Figure~\ref{fig:uef+acef} the
number of assets in the ACEF is below 10 for $R\geq 0.0068$.

\subsection{Variants of the problem}
\label{sec:extensions}

In this section, we briefly discuss other variants of the portfolio
selection problem and different constraint types not taken into
account in this work. 

\vspace{0.5cm}
\begin{description}
\item[Transaction lots:] \citeauthor{MaSp99} (\citeyear{MaSp99})
  consider the constraint stating that assets can be traded only in
  indivisible lots of fixed size.  In this case, the problem must be
  formulated in terms of integer-valued variables ---as opposed to
  real-valued ones--- that represent, for each asset, the number of
  purchased lots, instead of the real-valued ones.
  
  Given that assets are normally composed by units, this constraint is
  certainly meaningful; its practical importance however depends on the
  ratio between the size of the minimum trading lots and the size of
  the shares involved in the portfolio.
  
\item[Linear risk:] \citeauthor{KoYa91} (\citeyear{KoYa91}) propose to
  replace the objective function based on the variance with a linear
  risk function based on absolute deviation. This simplified model is
  easier to deal with due to the absence of the quadratic term.
  According to the authors, it provided results as accurate as in the
  basic model.  The effectiveness of the linear objective function,
  though, has been criticized by \citeauthor{Sima97}
  (\citeyear{Sima97}).
  
\item[Semi-variance:] \citeauthor{MTXY93} (\citeyear{MTXY93}) propose
  a different version of the objective function based on
  \emph{semi-variance} rather than on variance. Semi-variance is a
  concept similar to variance, except that it takes into account only
  downward deviations rather than upward and downward ones. In fact,
  as Markowitz (\citeyear{Mark59}, Cap.\ 9) first noted, an objective
  function based on variance has the drawback that it considers very
  high and very low returns equally undesirable.  The use of
  semi-variance instead tends to eliminate very low returns, without
  worrying about the distribution of positive returns.
  
\item[Skewness:] A more complex objective function is considered by
  Konno and coworkers \cite{KoSY93,KoSu95}, who introduce a
  third-order quantity called \emph{skewness}. Skewness is represented
  by means of an $n\times n\times n$ matrix, and contributes to the
  objective function with a cubic term. The authors also propose an
  algorithm to solve the resulting problem.
  
\item[Multi-period optimization:] Most of the literature refers to a
  one-period trade in which the investor has initially no asset
  shares. In the perspective of a longer run, however the investor at
  each round has to decide in which assets to invest, based also on
  the assets that he/she already holds.  \citeauthor{Yosh96}
  (\citeyear{Yosh96}) considers the case of the presence of an
  \emph{initial} portfolio and the notion of \emph{transaction costs}.
  The transaction costs must be subtracted from the expected return,
  and thus the number of atomic sell or buy operations must be taken
  into account.
\end{description}

\section{Local Search}
\label{sec:local-search}

Local search is a family of general-purpose techniques for search and
optimization problems.  Local search techniques are
\emph{non-exhau\-sti\-ve} in the sense that they do not guarantee to find
a feasible (or optimal) solution, but they search non-systematically
until a specific stop criterion is satisfied.

\subsection{Local Search Basics}

Given an instance $p$ of a problem $P$, we associate a \emph{search
  space} $S$ with it. Each element $s \in S$ corresponds to a
potential solution of $p$, and is called a \emph{state} of $p$.  An
element $s\in S$ that corresponds to a solution that satisfies all
constraints of $p$ is called a \emph{feasible state} of $p$.  Local
search relies on a function $N$, depending on the structure of $P$,
which assigns to each $s\in S$ its \emph{neighborhood} $N(s)\subseteq
S$.  Each $s'\in N(s)$ is called a \emph{neighbor} of $s$.

A local search algorithm starts from an initial state $s_0$, which can
be obtained with some other technique or generated randomly, and
enters a loop that \emph{navigates} the search space, stepping from
one state $s_i$ to one of its neighbors $s_{i+1}$.  
The neighborhood is usually composed by the states that are obtained
by some local change (called \emph{move}) from the current one.

We call $s_{final}$ the state reached after the stop condition has been met.
We write $s' = s \otimes m$ to mean that $s'$ is obtained from
$s$ applying the move $m$.

Local search techniques differ from each other according to the strategy
they use both to select the move in each state and to stop the search. In all
techniques, the search is driven by a \emph{cost function} $f$ that estimates
the quality of the state.

For search
problems, the cost function generally accounts for the number of
violated constraints, and thus the search is guided by the aim of
minimizing $f$ and reaching the value $0$ for it. For optimization
problems, $f$ also takes into account the objective function of the
problem. We call $s_{best}$ the first state such that $f(s_{best})
\leq f(s)$ for all $s$ encountered in the search started in $s_0$.
Depending on the search technique, $s_{best}$ may or may not coincide
necessarily with $s_{final}$.

In conclusions, local search consists in defining a search space and a
neighborhood relation, creating an initial state $s_0$, and moving
from $s_0$ to $s_{final}$ according to the chosen strategy.  

The most common local search techniques are \emph{hill climbing} (HC),
\emph{simulated annealing} (SA), and \emph{tabu search} (TS).  We describe
in more details TS which is the technique that gave best results for
our application.

\subsection{Tabu Search}
\label{sec:tabu-search}

Tabu search is a modern local search technique that has been
successfully applied in many real-life problems. A full description of
TS is out of the scope of this paper (see, e.g., \citeauthor{GlLa97},
\citeyear{GlLa97}). We now describe the formulation of the technique
which has been used in this work.

At each state $s_i$, TS explores exhaustively the current neighborhood
$N(s_i)$. Among the elements in $N(s_i)$, the one that gives the
minimum value of the cost function becomes the new current state
$s_{i+1}$, independently of the fact whether $f(s_{i+1})$ is less or
greater than $f(s_{i})$.

Such a choice allows the algorithm to \emph{escape} from local minima, but
creates the risk of cycling among a set of states.  In order to prevent
cycling, the so-called \emph{tabu list} is used, which determines the forbidden
moves. This list stores the most recently accepted moves. The \emph{inverses}
of the moves in the list are forbidden.

The simplest way to run the tabu list is as a queue of fixed size $k$.
That is, when a new move is added to the list, the oldest one is
discarded.  We employ a more general mechanism which assigns to each
move that enters the list a random number of moves, between two values
$k_{min}$ and $k_{max}$ (where $k_{min}$ and $k_{max}$ are parameters
of the method), that it should be kept in the tabu list.  When its
tabu period is expired, a move is removed from the list. This way the
size on the list is not fixed, but varies dynamically between
$k_{min}$ and $k_{max}$.

There is also a mechanism that overrides the tabu status: If a move
$m$ gives a ``large'' improvement of the cost function, then its
tabu status is dropped and the resulting state is acceptable as the
new current one.  More precisely, we define an \emph{aspiration
  function} $A$ that, for each value $v$ of the objective function,
returns another value $v'$ for it, which represents the value that the
algorithm ``aspires'' to reach from $v$.  Given a current state $s_i$,
the cost function $f$, and a neighbor state $s'$ obtained through the
move $m$, if $f(s') \leq A(f(s_i))$ then $s'$ can be accepted as
$s_{i+1}$, even if $m$ is a tabu move.

The stop criterion is based on the so-called \emph{idle iterations}:
the search terminates when it reaches a given number of iterations
elapsed from the last improvement of the current best state.

\subsection{Composite local search}

One of the attractive properties of the local search framework is that
different techniques can be combined and alternated to give rise to
complex algorithms.

In particular, we explore what we call the \emph{token-ring} strategy, which is
a simple mechanism for combining different local search techniques and/or
different neighborhood relations. Given an initial state $s_0$ and a set of
basic local search techniques $t_1,\dots,t_{q}$, that we call \emph{runners},
the token-ring search makes circularly a run of each $t_i$, always starting
from the best solution found by the previous runner $t_{i-1}$ (or $t_q$ if $i
= 1$).

The full token-ring run stops when it performs a fixed number of
rounds without an improvement by any of the solvers, whereas the
component runners $t_i$ stop according to their specific criteria.

The effectiveness of token-ring search for $2$ runners has been
stressed by several authors (see \citeauthor{GlLa97},
\citeyear{GlLa97}). In particular, when one of the two runners, say
$t_2$, is not used with the aim of improving the cost function, but
rather for diversifying the search region, this idea falls under the
name of \emph{iterated} local search (see, e.g., \citeauthor{Stut99},
\citeyear{Stut99}). In this case the run with $t_2$ is normally called
the \emph{mutation} operator or the \emph{kick} move.  

For example, we used the alternation of a TS using a small
neighborhood with HC using a larger neighborhood for the high-school
timetabling problem \cite{Scha99c}.

\section{Portfolio Selection by Local Search}
\label{sec:application}

In order to apply local search techniques to portfolio selection we
need to define the search space, the neighborhood structures, the cost
function, and the selection rule for the initial state.

\subsection{Search space and neighborhood relations}
\label{sec:search-space}

For representing a state, we make use of two sequences $L =
\{a_{l_1},$ $\dots,a_{l_p}\}$ and $S = \{x_{l_1},\dots,x_{l_p}\}$ such
that $a_{l_i}\in A$ and $x_{l_i}$ is the fraction of $a_{l_i}$ in the
portfolio.  All assets $a_j\not\in L$ have the fraction $x_j$
implicitly set to $0$.  With respect to the mathematical formulation,
having $a_i\in L$ corresponds to setting $z_i$ to 1.

We enforce that the length $p$ of the sequence $L$ is such that $p\leq
k$, that the sum of $x_{l_i}$ equals $1$, and that $\epsilon_{l_i}
\leq x_{l_i} \leq \delta_{l_i}$ for all elements in $L$. Therefore,
all elements of the search space satisfy 
Constraints~(\ref{con:sum-one}),~(\ref{con:limit}),~(\ref{con:card}), and
(\ref{con:mm_limit}). Constraint~(\ref{con:return}) instead is not always
satisfied and it is included in the cost function as explained below.

\vspace{\baselineskip}

Given that the problem variables are continuous, the definition of
the neighborhood relations refers to the notion of the \emph{step} of a
move $m$, which is a real-valued parameter $q$, with $0 < q < 1$, that
determines the quantity of the move. Given a step $q$,
we define the following three neighborhood relations:
Other relations have been investigated, but did not provide valuable
results.

\begin{small}
\begin{description} 
\item[\underline{\idr{}}] (\textsf{[i]}ncrease, \textsf{[d]}ecrease,
  \textsf{[R]}eplace):
  \begin{description} 
  \item[Description:] The quantity of a chosen asset is increased or
    decreased. All other shares are changed accordingly so as to
    maintain the feasibility of the portfolio. If the share of the
    asset falls below the minimum it is replaced by a new one.
  \item[Attributes:] $\langle a_i,s,a_j\rangle$ with $a_i \in A$, $s \in \{\uparrow,\downarrow\}$, $a_j \in A$
  \item[Preconditions:] $a_i\in L$ and $a_j \not\in L$
  \item[Effects:] If $s =\:\uparrow$ then $x_i := x_i\cdot(1+q)$,
    otherwise $x_i := x_i\cdot(1-q)$. All values $x_k-\epsilon_k$ are
    renormalized so as to maintain the property that $x_k$'s add up to
    1.  We renormalize $x_k-\epsilon_k$ and not $x_k$ to ensure that
    no asset rather then $a_i$ can fall below the minimum.
  \item[Special cases:] If $s=\:\downarrow$ and $x_i (1-q) < \epsilon_i$,
    then $a_i$ is deleted from $L$ and $a_j$ is inserted with $x_j =
    \epsilon_j$.  If $s=\:\uparrow$ and $x_i (1+q) > \delta_i$, then $x_i$
    is set to $\delta_i$.
  \item[Reference:] Revised version of 
    \citeauthor{CMBS00}\ (\citeyear{CMBS00}).
  \end{description}  
\item[\underline{\idid{}}] (\textsf{[i]}ncrease, \textsf{[d]}ecrease, \textsf{[I]}nsert, \textsf{[D]}elete):
  \begin{description} 
  \item[Description:] Similar to \idr{}, except that the deleted asset
    is not replaced and insertions of new assets are also considered.
  \item[Attributes:] $\langle a_i,s\rangle$ with $a_i \in A$, $s\in \{\uparrow,\downarrow,\hookrightarrow\}$
  \item[Preconditions:] If $s =\:\downarrow$ or $\uparrow$ then $a_i\in L$. If
    $s =\:\hookrightarrow$ then $a_i \notin L$.
  \item[Effects:] If $s =\:\uparrow$ then $x_i := x_i\cdot(1+q)$;
    if $s =\:\downarrow$ then $x_i := x_i\cdot(1-q)$; if $s =\:\hookrightarrow$
    then $a_i:\epsilon_i$ is inserted into $L$. The portfolio is
    repaired as explained above for \idr{}.
  \item[Special cases:] If $s=\:\downarrow$ and $x_i (1-q) < \epsilon_i$,
    then $a_i$ is deleted from $L$, and it is not replaced. If
    $s=\:\uparrow$ and $x_i (1+q) > \delta_i$, then $x_i$ is set to
    $\delta_i$.
  \end{description}
  
\item[\underline{\tid{}}] (\textsf{[T]}ransfer, \textsf{[I]}nsert, \textsf{[D]}elete):
  \begin{description} 
  \item[Description:] A part of the share is transferred from one asset
    to another one. The transfer can go also toward an asset not in the
    portfolio, which is then inserted. If one asset falls below the
    minimum it is deleted.
  \item[Attributes:]  $\langle a_i,a_j\rangle$ with $a_i\in A$, $a_j\in A$
  \item[Preconditions:] $a_i\in L$
  \item[Effects:] The share $x_i$ of asset $a_i$ is decreased by $q
    \cdot x_i$ and $x_j$ is increased by the same quantity.  If
    $a_j\not\in L$ than it is inserted in $L$ with the quantity $q
    \cdot x_i$.
  \item[Special cases:] The quantity transferred is larger than $q
    \cdot x_i$ in the following two cases: $(i)$ If after the decrease
    of $x_i$ we have that $x_i < \epsilon_i$ then also the remaining
    part of $x_i$ is transferred to $a_j$.  $(ii)$ If $a_j\not\in L$
    and $q \cdot x_i < \epsilon_j$ then the quantity transferred is
    set to $\epsilon_j$.
  \item[Reference:] Extended version of \citeauthor{Roll97} (\citeyear{Roll97}).
  \end{description}
\end{description}
\end{small}

Intuitively, \idr{} and \idid{} increase (or decrease) the quantity of
a single asset, whereas \tid{} trasfers a given amount from one asset
to another one.

Notice that \idr{} moves never change the number of assets in the
portfolio, and thus the search space is not connected under \idr{}.
Therefore, the use of \idr{} for the solution of the problem is
limited.  The relation \idid{} in fact is a variant of \idr{} that overcomes
this drawback.

Notice also that under all three relations the size of the
neighborhood is not fixed, w.r.t.\ the size of $L$, but it depends on
the state. In particular, it depends on the number of assets that
would fall below the minimum in case of a move that reduces the quantity of
that asset.  For example, for \idr{}, the size is linear,
$2\cdot\card{L}$, if no asset $a_i$ is such that $x_i(1-q) <
\epsilon_i$, but becomes quadratic, $\card{L} + \card{L} \cdot (n -
\card{L})$, if all assets are in such conditions.

We now define the inverse relations, which determines which moves are
tabu. Our definitions are the following: For \idr{} and \idid{}, the
inverse of $m$ is any move with the same first asset and different
arrow. For \tid{}, it is the move with the two assets exchanged.

\subsection{Cost function and initial state}
\label{sec:cost-function}

Recalling that all constraints but Constraint~(\ref{con:return}) are
automatically satisfied by all elements of the search space, the cost
function $f(X)$ is composed by the objective function and the degree
of violation of Constraint~(\ref{con:return}). Specifically, we define
two components, $f_1(X) = max(0,\sum_{i=1}^{n} r_i x_i - R)$ and
$f_2(X) = \sum_{i=1}^{n}\sum_{j=1}^{n} \sigma_{ij} x_{i} x_{j}$, which
take into account the constraint and the objective function,
respectively.  The overall cost function is a linear combination of
them: $f(X) = w_1 f_1(X) + w_2 f_2(X)$.

In order to ensure that a feasible solution is found, $w_1$ is
(initially) set to a much larger value than $w_2$. However, during the
search, $w_1$ is let to vary according to the so-called \emph{shifting
  penalty} mechanism (see, e.g., \citeauthor{GeHL94},
\citeyear{GeHL94}): If for $K$ consecutive iterations
Constraint~(\ref{con:return}) is satisfied, $w_1$ is divided by a
factor $\gamma$ randomly chosen between 1.5 and 2.  Conversely, if it
is violated for $H$ consecutive iterations, the corresponding weight
is multiplied by a random factor in the same range (where $H$ and $K$
are parameters of the algorithm).

Notice that given a state $s$ and a move $m$ the evaluation of the
cost change associated to $m$, i.e. $f(s\otimes m) - f(s)$ is computationally
quite expensive for both \idr{} and \idid{}, due to the fact
that $m$ changes the fraction of all assets in $L$.  The computation
of the cost is instead much cheaper for \tid{}.


The initial state is selected as the best among $I=100$ random
portfolios with $k$ assets.  However, experiments show that the
results are insensitive to $I$.

\subsection{Local search techniques}

We implemented all the three basic techniques, namely HC, SA, and TS,
for all neighborhood relations.  HC, which performs only improving and
sideways moves, is implemented both using a random move selection and
searching for the best move at each iteration (steepest descent).  SA,
which for the sake of brevity is not described in this paper, is
implemented in the ``standard'' way described in \cite{JAMS89}.
TS is implemented as described in Section~\ref{sec:tabu-search} 
using a tabu list of  variable size and the shifting penalty mechanism.

We also implemented several token-ring procedures. The main idea is to
use one technique $t_1$, with a large step $q$, in conjunction with
another one $t_2$, with a smaller step. The 
technique $t_1$ guarantees diversification, whereas $t_2$ provides
a ``finer-grain'' intensification.  

The step $q$ is not kept fixed for the entire run, but it is allowed
to vary according to a random distribution.  Specifically, we
introduce a further parameter $d$ and for each iteration the step is
selected with equal distribution in the interval $q-d$ and $q+d$.

Due to its limited exploration capabilities, \idr{} is used only for
$t_2$. Other combinations, of two or three techniques, have
also been tested as described in the experimental results.

\subsection{Benchmarks and experimental setting}

We experiment our techniques on 5 instances taken from real stock
markets.\footnote{Available at the URL
  \texttt{http://mscmga.ms.ai.ac.uk/jeb/or\-lib/port\-fo\-lio.html}}
We solve each instance for 100 equally distributed values for the
expected return $R$.

We set the constraint parameters exactly as \citeauthor{CMBS00}
(\citeyear{CMBS00}): $\epsilon_i = 0.1$ and $\delta_i = 1$ for
$i=1,\dots,n$, and $k = 10$ for all instances.

Given that the constraint problem has never been solved exactly, we
cannot provide an absolute evaluation of our results.  
We measure the quality of our solutions in average percentage loss
w.r.t.\ the UEF (available from the web site).  We also refer to the
ACEF, which we obtain by getting, for each point, the best solution
found by the set of all runs using all techniques.  The ACEF has been
computed using a very large set of long runs, and reasonably provides a good
approximation of the optimal solution of the constrained problem.

Table~\ref{tab:instances} illustrates, for all instances, the original
market, the average variance of UEF (multiplied by $\times 10^3$ for
convenience), and the percentage average of the difference between
ACEF and UEF.

\begin{table}[tbhp]
\begin{center}
\begin{tabular}[tbhp]{|l|l|c||c|c|}\hline
No.\ & Origin & assets & UEF &  \% Diff   \\ \hline\hline 
1 & Hong Kong & 31 & 1.55936  &  0.00344745\\
2 & Germany & 85 & 0.412213  &  2.53845\\    
3 & UK  & 89 & 0.454259  & 1.92711 \\        
4 & USA  & 98 & 0.502038 & 4.69426  \\       
5 & Japan & 225 & 0.458285 & 0.204786 \\ \hline
\end{tabular}
\end{center}
\caption{The benchmark instances}\label{tab:instances}
\end{table}

Notice that the problem for which the discrepancy between UEF and ACEF
is highest is no.\ 4 (with 4.69\%). For this reason we illustrate our
results for no.\ 4, in which the differences are more tangible. 

Except for no.\ 1, all other instances give qualitatively similar
results and they require almost the same parameter settings. Instance
no.\ 1 instead, whose size is considerably smaller than the others,
shows peculiar behaviors and requires completely different settings.
Specifically, it requires shorter tabu list and much smaller steps.

\section{Experimental Results}
\label{sec:experimental-results}

The code is written in C++ with the compiler \texttt{gcc} (version
2.96), and makes use of the local search library \local{}
\cite{DiSc00}. It runs on a 300MHz Pentium II using Linux.

In the following experiments, we run 4 trials for every point.  For
each parameter setting, we therefore run 2000 trials (4 trials
$\times$ 100 points $\times$ 5 instances).  Except for the first point
of the UEF, in one of the four trials the initial state is not random,
but it is the final state of the previous solved point of the UEF.
The number of iterations is chosen in such a way that each single
trial takes approximately 2 seconds (on a 300MHz Pentium II, using the
C++ compiler egcs-2.91.66), and therefore each test runs for
approximately an hour.

We experimented with 20 different values of the step $q$.  Regarding
the step variability $d$, preliminary experiments show that the best value is
$q$, which means the step varies between $0$ and $2q$. In all the
following experiments, $d$ is either set to $0$ (fixed step) or is set
to $q$ (random step).

Regarding the parameters related to the shifting penalty mechanism,
the experiments show that the performances are quite insensitive to
their variations as far as they are in a given interval. Therefore, we
set such parameters to fixed values throughout all our experiments ($H
= 1$, $K = 20$).

\subsection{Single solvers}

The first set of experiments regards a comparison of algorithms
using the three neighborhood relations \idid{}, \idr{}, and \tid{}
in isolation. Given that the search space is not connected
under \idr{}, the relation \idr{} is run for initial states of all
sizes from $2$ to $10$ (it is therefore granted a much longer running
time).  For the other two, \idid{} and \tid{}, we start always with an
initial state of $10$ assets.

Table~\ref{tab:compare-nhe} shows the best results for TS for both fixed and
random steps, and the corresponding step values.  For TS, the tabu
list length is 10-25, and the maximum number of idle iterations is set
to 1000.  

The table shows also the best performance of HC and SA for \tid{} and
\idid{}. For the sake of fairness, we must say that the parameter
setting of SA has not been investigated enough.

The results in Table~\ref{tab:compare-nhe} show that TS works much
better than the others, and \tid{} works better than \idr{} and
\idid{}.  They also show that the randomization of the step improves
the results significantly.

\begin{table}[tbhp]
\begin{center}
\begin{tabular}[tbhp]{|cc||cc|cc|}\hline
&  & \multicolumn{2}{|c}{Fixed Step} &   
                \multicolumn{2}{|c|}{Random Step}\\ 
Tech.~& ~ Nhb~ & ~ step ~ & \% Diff & ~base step~ & \% Diff \\ \hline
TS& \idid  & 0.5   & 6.31568 & 0.4  &  5.60209    \\
TS& \tid   & 0.5   & 5.42842 & 0.3  &  4.85423    \\
TS& \idr   & 0.4   & 5.4743  & 0.4  & 5.4621      \\
SA& \tid   & 0.4   & 53.7006 & 0.4  &  56.5798           \\
SA& \idid  & 0.2      & 118.698        & 0.5  &  113.735           \\
HC& \tid    & 0.2  & 29.2577 & 0.2   & 29.039    \\ 
HC& \idid  & 0.2  & 41.4734   & 0.1   & 41.0438    \\ \hline
\end{tabular}
\caption{Comparison of simple solvers}\label{tab:compare-nhe}
\end{center}
\end{table}

\subsection{Composite solvers}

Table~\ref{tab:compare-tokenring} shows the best results for
token-ring with various combinations of two or three neighborhoods all
using TS and random steps.  Notice that we consider as token-ring
solver also the interleaving of the same technique with different
steps.

\begin{table}[tbhp]
\begin{center}
\begin{tabular}[tbhp]{|cc|cc|cc||c|}\hline
\multicolumn{2}{|c|}{Runner 1} & \multicolumn{2}{c|}{Runner 2}& 
\multicolumn{2}{c||}{Runner 3} &  \\
Nbh & Step & Nbh & Step & Nbh & Step & \% Diff \\ \hline
\tid & 0.4 & \tid & 0.05 & - & - &  4.70872\\
\tid & 0.4 & \tid & 0.04 & \tid & 0.004 & 4.70866\\    
\tid & 0.4 & \idr & 0.05 & - & - & 4.70804\\
\tid & 0.4 & \idr & 0.05 & \tid & 0.01 & 4.71221\\
\idid & 0.4 & \idid & 0.04 & - & - & 5.06909\\
\idid & 0.3 & \idid & 0.03 & \idid & 0.003 & 4.99406\\
\idid & 0.4 & \idr & 0.05 & - & - & 4.99206 \\
\idid & 0.4 & \idr & 0.04 & \idid & 0.004 & 5.16368\\ \hline
\end{tabular}
\end{center}
\caption{Comparison of composite solvers}\label{tab:compare-tokenring}
\end{table}

The table shows that the best results are obtained using the
combination of \tid{} and \idr{}, but \tid{} with different steps
performs almost as good. This results are very close to the ACEF
(4.69426\%), which is obtained using also much longer runs (24 hours each).

In conclusion, the best results (around 4.7\%) are obtained by token
ring solvers with random steps.  Further experiments show that the
most critical parameter is the size of the step of $t_1$, which must
be in the range [0.3, 0.6]. They also show that using alternation of
fixed steps the best result obtained is 4.84883.

\subsection{Effects of constraints on the results}

\begin{figure}[htbp]
  \begin{center}
    \includegraphics[width=60mm,height=90mm,angle=-90]{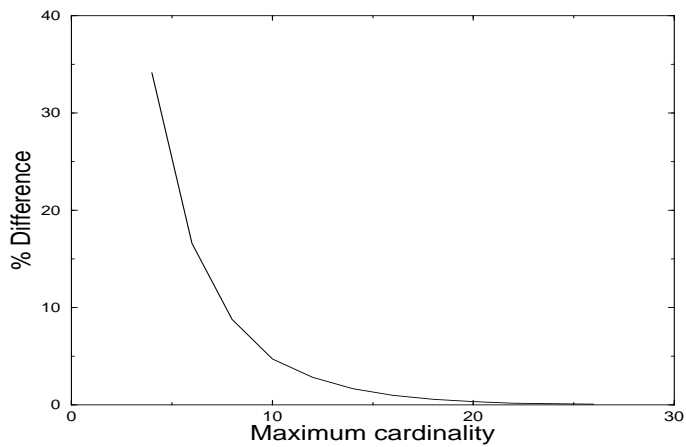}
    \caption{Results for different values of the maximum cardinality ($k$)}
    \label{fig:card}
  \end{center}
\end{figure}

We conclude with a set of experiments that highlights the role played
by constraints~\ref{con:card} and~\ref{con:mm_limit} on the problem.
Figure~\ref{fig:card} shows the best results for instance no.\ 4 for
different values of the maximum number of assets $k$ ($\epsilon_i$ and
$\delta_i$ are fixed to the values $0.01$ and $1$).

The results show that the effect of the constraint decreases quite
steeply when increasing $k$. The effect is negligible for $k >
30$.

\begin{figure}[htbp]
  \begin{center}
    \includegraphics[width=60mm,height=90mm,angle=-90]{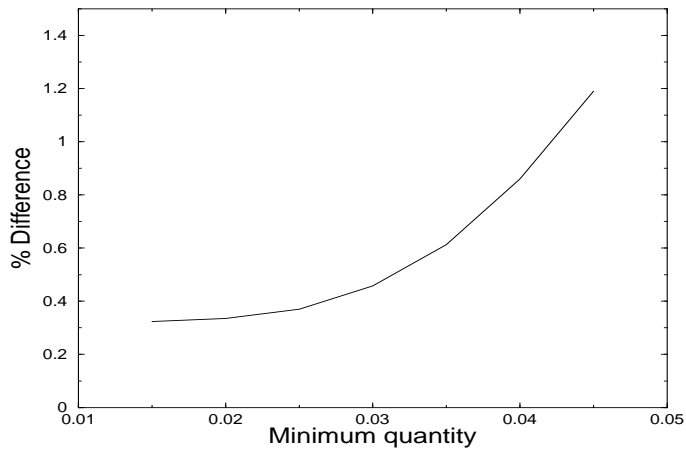}
    \caption{Results for different values of the minimum quantity ($\epsilon_i$)}
    \label{fig:mm_limit}
  \end{center}
\end{figure}

Figure~\ref{fig:mm_limit}, instead, shows how the quality of the portfolio
decreases while increasing the minimum quantity ($\epsilon_i$).  In order to
focus on the minimum quantity constraint, we use a high value
for the maximum cardinality ($k=20$) so as to make the effect of the
corresponding cardinality constraint less visible.

We don't show the results for different values of $\delta_i$ because
the constraint on maximum quantity is less important from the
practical point of view.

\section{Related Work}
\label{sec:related-work}

This problem has been previously considered by \citeauthor{CMBS00}
(\citeyear{CMBS00}), who implemented three solvers based on TS, SA,
and genetic algorithms (GA). Their experimental results suggest that
GA and SA work better than TS.  Even though the TS procedure is not
completely explained in the paper, we believe that this ``defeat'' of
TS in favor of SA and GA is due to the fact that their version of
TS is not sufficiently optimized.

The neighborhood relation used by \citeauthor{CMBS00} is a variant of
\idr{}. The difference stems from the fact that in their case a move
$m$ is represented by only the pair $\langle a_i,s\rangle$ and the
replacing asset $a_j$ is not considered part of $m$, but it is
randomly generated whenever necessary. This definition makes
incomplete the exploration of the full neighborhood because the
quality of a move $\langle a_i,\downarrow\rangle$ may depends of the
randomly generated $a_j$. In our work, instead, all possible replaces
$a_j$ are analyzed.  In addition, the application of a move $m=\langle
a_i,\downarrow\rangle$ is \emph{non-deterministic}, and therefore it
is not clear which is the definition of the inverse of $m$, and the
definition of the tabu mechanism.  Finally, with respect to our
version, their TS misses the following important features: shifting
penalty mechanism, random step (they use the fixed value $0.1$), and
variable-size tabu list.

Even though \citeauthor{CMBS00}\ solve the same problem instances,
a fair comparison between their and our results is not
possible for two reasons:

First, they formulate Constraint~(\ref{con:card}) with the equality
sign, i.e.\ $\sum_{i=1}^{n} z_i = k$, rather than as an inequality.
As the authors themselves admit, constraining the solution to an
exact number of assets in the portfolio is not meaningful by itself,
but it is a tool to solve the inequality case.
They claim that the solution of the problem with the inequality can be
found solving their problem for all values from $1$ to $k$.
Unfortunately, though, they provide results only for the problem with
equality.

Second, they do not solve a different instance for each value of $R$,
but (following \citeauthor{Pero84}, \citeyear{Pero84}), they
reformulate the problem without Constraint~(\ref{con:return}) and with
the following objective function: $f(X) = \lambda f_1(X) +
(1-\lambda)f_2(X)$.  The problem is then solved for different values
of $\lambda$.  The quality of each solution is measured not based on
the risk difference w.r.t.\ the UEF for the same return $R$, but using
a metric that takes into account the distance to both axis.  The
disadvantage of this approach is that they obtain the solution for
a set of values for $R$ which are not an homogeneously distributed.
Therefore their quality cannot be measured objectively, but it depends
on how much they cluster toward the region in which the influence of
Constraints~(\ref{con:card}) and~(\ref{con:mm_limit}) is less or more
strong. In addition, these sets of points are not provided, and thus
the results are not reproducible and not comparable.

\citeauthor{Roll97} (\citeyear{Roll97}) considers the unconstrained
problem therefore his results are not comparable.  
He introduces the \tid{} neighborhood which turned out to be the most
effective. Although, the definition of \citeauthor{Roll97} is
different because he considers only transfers and no insertions and
deletions.  This is because, for the unconstrained problem, all assets
can be present in the portfolio at any quantity, and therefore there
is no need of inserting and deleting.  The introduction of insert and
delete moves is our way to adapt his (successful) idea to the constrained case.

Rolland makes use of a tabu list of fixed length equal to $0.4\cdot
n$, thus linearly related to the number of assets.  He alternates the
fixed step value 0.01 with the fixed value 0.001, shifting every 100
moves.  Our experiments confirm the need for two (and no more than
two) step values, but they show that those values are too small for
the constrained case. In addition, for the constrained problem,
randomization works better than alternating two fixed values.

\section{Conclusions and Future Work}
\label{sec:conclusions}

We compared and combined different neighborhood relations and local
search strategies to solve a version of the portfolio selection
problem which involves a mixed-integer quadratic problem.  Rather than
exploring all techniques in the same depth, we focussed on TS that
turned out to be the most promising from the beginning.

This work shows also how adaptive adjustments and randomization could
help in reducing the burden of parameter setting.  For example, the
choice of the step parameter turned out to be particularly critical.

We solved public benchmark problems, but unfortunately no comparison
with other results is possible at this stage.

In the future, we plan to adapt the current algorithms to different
versions of the portfolio selection problem, both discrete and
continuous, and to related problems. Possible hybridization of local
search with other search paradigms, such as genetic algorithms, will
also be investigated.

\section*{Acknowledgements}

I thank T.-J. Chang and E. Rolland for answering all questions about
their work, and K. R. Apt and M. Cadoli for comments on earlier drafts
of this paper.

\bibliographystyle{named}
\bibliography{medium_string,portfolio,timetabling}

\end{document}